\documentclass[aps,twocolumn,10pt]{revtex4}  
\usepackage{graphicx}
\usepackage[dvips]{color}
\input{epsf.tex}
\newcommand{\bibi}{\bibitem}                                                  
\newcommand{\etal}{\it {et al.}}                                             

\newcommand{\half}{\frac {1}{2}}                                             
                                             
\newcommand{\beq}{\begin{equation}}                                           
\newcommand{\eeq}{\end{equation}\noindent}                                  
\newcommand{\beqr}{\begin{eqnarray}}                                          
\newcommand{\eeqr}{\end{eqnarray}\noindent}                                   
\newcommand{\vQ}{\bf Q}                                                      
\newcommand{\vcr}{{\bf r}}                                                     
\newcommand{\vk}{{\bf k}}                                                     
                                                     
\newcommand{\vq}{{\bf q}}

\newcommand{\cd}{c^{\dag}} 
\newcommand{\bd}{b^{\dag}} 
\newcommand{\hd}{h^{\dag}}

\begin{document} 
\title{Charge pair hopping and Bose-Einstein condensation in underdoped Mott insulators}
\author{Sanjoy K. Sarker and Timothy Lovorn\\                                                   
Department of Physics and Astronomy \\                                        
The University of Alabama, Tuscaloosa, AL 35487 \\}                           
\begin{abstract}
Recently, a renormalized Hamiltonian has been derived by continuing spin states from the Mott 
limit of the $t$-$J$ model to the underdoped region. We show that it
naturally leads to a pairing mechanism in which the pair has a dual character. Its spin part is a spinon 
singlet which condenses at $T^*$. The charge part is a real-space holon pair formed at $T_p < T^*$, which 
undergoes Bose-Einstein condensation at $T_c < T_p$. While neither is
observable separately, the combination is. The mechanism is consistent with the small superfluid density,
the decline of $T_c$ at small doping, and the existence of pairs above $T_c$ in cuprates, 
indicated by the observation of diamagnetism and Nernst effect.

   
\end{abstract} 
\maketitle                                                               
\vspace{0.5 in}                                                              



In a conventional superconductor the metallic state is characterized by
quasielectrons, which form spatially overlapping Cooper pairs. 
In contrast, the elementary excitations in the metal phase
of a hole-doped cuprate superconductor are hidden, and their nature
unknown, which raises the questions: what is paired and how? 
Here we address this issue using the $t$-$J$ model, which describes 
electrons of concentration $1 - x$ hopping on a lattice such that no site 
is doubly occupied. At half filling ($x = 0$), the system is a
Mott insulator, with electrons localized as moments (or spins) which 
interact antiferromagnetically. Unexpectedly, the metallic state in cuprates 
is two dimensional (2d), even though these are (layered) 3d materials. 
Anderson argued that in 2d electrons remain localized even for $x > 0$, 
metallic conduction results from the motion of positively charged 
spinless holons. It is then natural to invoke continuity and construct a theory
of the doped region by continuing the spin states from the insulator \cite{and}.
Experimental support comes from the fact that the carrier concentration is $x$, 
not $1-x$. Also, despite the reappearance of nodal quasielectrons 
below $T_c$, the $T = 0$ superfluid density $\rho_s(0) \sim x$ \cite{orn1}, which is 
suggestive of holon pairing. 

In the $t$-$J$ model, hopping by a localized electron can be described as 
an exchange of a spin-1/2 particle (spinon) and a spinless holon, and the AF 
interaction as an exchange of two spinons. This reflects the $U(1)$ 
gauge symmetry of the model. The electron field is then represented as 
$\cd _{i\sigma} = \bd _{i\sigma}h_i$, where $\bd _{i\sigma}$ creates a spinon of spin 
$\sigma$ and $h_i$ destroys a holon, such that the particle number at each site $i$ is 
conserved and equals 1. The ground state at half filling is known: it is a mixture of 
the Arovas-Auerbach valence-bond (VB) state, in which bosonic spinons are paired 
into singlets, and the Neel state \cite{aro,sar1}. The VB state survives the destruction of 
AF order up to a temperature $T^*$. However, despite many attempts, the connection
with the doped phase (with fermionic holons) has not been established theoretically.

Recently, we have solved the continuation problem and derived a renormalized 
Hamiltonian valid for small $x$, assuming that hole
motion prevents long-range magnetic order beyond some small $x_c$ \cite{sar9,sar10}. This implies
that spin excitations are gapped since spinons are bosons.
The new Hamiltonian has the symmetry of the original model, which
allows us to determine its phases by  continuing all the contiguous spin
phases from half-filling. The approach is blind:  no experimental inputs are 
introduced by hand. Thus, we avoid the universal practice of assuming
two dimensionality of the normal state; instead it emerges as a consequence
of the theory \cite{sar10}. This is fundamental since 2d confinement is thought to be 
responsible for the unusual behavior. It provides a serious test  
since it is almost impossible to confine metallic conduction in the 
presence of 3d hopping; it has not been demonstrated in previous theories.  

Initially no new order is introduced, so that the phases are fully 
constrained by the symmetries at half filling. Yet, we obtain 
exactly two normal phases for small $x$ as in cuprates  \cite{tim}, with properties that match 
experiments, but disagrees with other theories (see, \cite{sar10} for details). 
(1) Recent experiments have
shown that, for small $x$, the high-$T$ strange-metal phase actually behaves 
like an insulator, with no Drude peak and a resistivity exceeding the
Mott maximum, consistent with dynamically localized charge carriers \cite{tak}.
In our case, since the high-$T$ phase has the
symmetry of the Hamiltonian, it is automatically an
insulator, as holons are localized by gauge forces.
(2) For $T < T^*$, the pseudogap temperature, holons form a spinless Fermi liquid of concentration $x$,
in agreement with recent experiments \cite{tak,ando,pad,tai}.
(3) The renormalized Hamiltonian has a pair hopping term, 
which automatically leads to $d$-wave superconductivity.
(4) There is strong evidence for a spin gap \cite{tim} which causes a downturn in the
paramagnetic susceptibility below $T^0 > T^*$.

Here we take a closer look at pairing. We show that ours
is {\em necessarily} a strong-coupling Hamiltonian which, together
with low density, induces holons to form real space pairs below $T_{p}$.
These undergo a Bose-Einstein condensation (BEC)\cite{mohit} below $T_c < T_p$,
which is qualitatively consistent with the observed downturn of $T_c$ at small $x$, 
and diamagnetism \cite{ong1} and Nernst effect \cite{ong2} above $T_c$. A key
prediction is a composite gauge-invariant {\lq\lq Cooper pair}", which is observable
above and below $T_c$. It is a well-defined excitation of momentum $\vq$ and energy
$\omega (\vq)$, whose charge is carried by a mobile (bound) holon pair, which picks up the 
local phase of spinon singlets (representing localized electrons) already condensed 
below $T^*$. 

The $t$-$J$ Hamiltonian for a layered 3d system is given by
\beq H = - \sum _{ij,\sigma} t_{ij} \cd _{i\sigma}c_{j\sigma} ~-~ 
  2\sum _{ij}J_{ij} {\rm A}^{\dag}_{ij}{\rm A}_{ij}.  \eeq
The first term describes electron hopping from $j$ to $i$ 
such that no site is doubly occupied, the second is the 
exchange interaction between spins. Here,
${\rm A}_{ij} = \half \lbrack b_{i\uparrow}b_{j\downarrow} - 
b_{i\downarrow}b_{j\uparrow}\rbrack$ destroys a spinon singlet,
and $J_{ij} =  4t_{ij}^2/U$, where $U$ is the Hubbard repulsion. 
Minimally, we need nearest-neighbor ($t$) and next-nearest-neighbor 
hopping ($t^{\prime}$) within each plane, and nearest-neighbor out-of-plane 
hopping $t_z << t$. For cuprates, $t/J \sim 3-4$. In general,
$t^{\prime}/t$ is small, so is $J^{\prime} = (t^{\prime}/t)^2J$, and will be 
neglected. 

As described in \cite{sar10}, renormalization proceeds in two steps. 
Hole motion is violently opposed by AF correlations, which localizes the hole in a small 
region, within which it hops rapidly.
The renormalized hole - bare hole plus surrounding spins - moves slowly through the lattice,
preventing AF order. By continuity, the Hamiltonian 
has the same form as the original one,
with renormalized parameters $t_{eff}$, $J_{eff}$ etc, except that spinons
have a gap $\Delta _s$. The spin gap is the singlet
breaking scale $\Omega \sim 2\Delta _s$. At small $x$, we expect $J_{eff}$,
$\Omega$ to scale with $J$. One-hole calculations suggest that for $t > J$, 
in-plane hopping amplitude $t_{eff} < J$, and also  scales with $J$.
The spin gap allows us to decouple unpaired spinons from hopping, and
derive a minimal Hamiltonian involving {\em sublattice-preserving}
hopping of renormalized holons and singlets only.
Now, due to the gap, singlets form within the plane since 
$J_z/J = (t_z/t)^2 << 1$ and a spinon belongs to one singlet at a time.
As a renormalized hole hops it breaks a singlet. For hops within the plane,
the excess energy $\Omega$ is removed if the hole makes a second hop, and
the singlet is reconstructed. Eliminating the intermediate state perturbatively we obtain 
\cite{sar10}
\beq H_1 = -  \frac{t_s}{2}(1-x)\sum _{ijl} {\rm A}^{\dag}_{jl}{\rm A}_{ij}h^{\dag}_ih_l, \eeq
which describes nearest- and next-nearest-neighbor hopping 
{\em within the same sublattice of the plane}, accompanied by a singlet backflow. 
The $(1-x)$ factor reflects the absence of a hole in the intermediate state. 
For a hop to a neighboring plane, two unpaired spins are created, one in
each plane. Now the system can not relax by the hole hopping to a third plane since
that would create more unpaired spins, which proliferate exponentially, blocking
hole motion in the $z$ direction. 

The system can also relax if a second hole follows the first, and the singlet is reconstructed.
The net effect is hopping by a holon pair to a parallel link, accompanied by a singlet backflow.
This mechanism also works for interplane hopping, since the entire singlet is transferred. 
Eliminating the intermediate state, we get
\beq H_2 = -t_s\sum _{ijlm;z} {\cal C}^{\dag}_{ml}(z){\cal C}_{ij}(z) 
- t_{sz}\sum _{ij;z} {\cal C}^{\dag}_{ij}(z){\cal C}_{ij}(z+1) + h.c., \eeq
where $t_{sz} = t_{z,eff}^2/4\Omega$, and  
${\cal C}_{ij} = 
(c_{j\downarrow}c_{i\uparrow} - c_{j\uparrow}c_{i\downarrow})/2 = -{\rm A}_{ij}{\rm F}^{\dag}_{ij}$
destroys a singlet made from physical electrons, which
is equivalent to destroying a spinon singlet and creating a holon pair with 
${\rm F}^{\dag}_{ij} = \hd _i\hd _j$. The first term describes intraplane, and the second, interplane 
pair hopping. The full Hamiltonian is gauge invariant and has the symmetries of the 
original model, plus an additional symmetry: total number of holes in each sublattice is conserved. 

Since the high-$T$ phase has full symmetry (nonorderd) holons are localized by gauge forces.
A pseudogap metal appears below $T^*$ as the spinon singlets condense, allowing
holons to propagate coherently. It is connected to the VB state at $x = 0$,
and is characterized by the order parameter
\beq A_{ij} = <{\rm A_{ij}}> = Ae^{i\half\vQ.(\vcr_i-\vcr_j)},\eeq 
or, its gauge-related copies, where $\vQ$ is the two-sublattice wave vector. 
This state has a quantum lattice order: on average singlets connect spinons on opposite 
sublattices. If we replace $\rm A_{ij}$ by its average $A_{ij}$,
the single-hole term [Eq. 2] describes free holon hopping within the sublattice, with a spectrum
$\epsilon _h =  - 2t_h + 2t_h(\sin k_x + \sin k_y)^2 + 4D_1t_sA^2 \sin k_x\sin k_y$,
where $t_h = t_sA^2(1-x)$. The last term is the Hartree contribution from the
pair hopping term, and $D_1 < x$ is the average hole hopping amplitude. 
At low-$T$, the pseudogap phase is thus a spinless Fermi liquid of concentration $x$.
The corresponding small Fermi pockets are not gauge invariant, but have been seen
indirectly in de Hass-van Alphen type experiments \cite{tai}. 

The condensate part of the pair-hopping term becomes
\begin{eqnarray}
 H_{2h} & = & - t_sA^2 \sum _{ij,lm;z}\lbrack {\rm F}^{\dag}_{ij}(z){\rm F}_{ml}(z) + h.c.
\rbrack \nonumber \\ 
& & - t_{zs}A^2\sum _{ij,z}\lbrack 
{\rm F}^{\dag}_{ij}(z){\rm F}_{ij}(z+1)+ h.c \rbrack. 
\end{eqnarray}
It clearly leads to 3d superconductivity via pair 
condendensation, so that $F_{ij} \equiv <{\rm F}_{ij}> \ne 0$. The
electron pair amplitude is then $<{\cal C}_{ij}> = - <{\rm A}_{ij}><{\rm F}^{\dag}_{ij}> \ne 0$. 
A $T = 0$ MF analysis showed that it is a robust $d$-wave, essentially
due to the symmetry of the VB state at $x = 0$. 

The MF approximation clearly would not work for $T > 0$ since the intraplane pair-hopping 
energy scale $t_s$ is essentially the same as that for single-hole hopping, ($t_s(1-x)$)
(Eqs. 2, 3) because they arise from the same singlet breaking mechanism. It would induce holons to form 
real-space pairs, particularly since (a) at small $x$, pairs would not overlap, and (b) they can 
reduce energy further by delocalizing in the $z$-direction.
Here we show this by analyzing pair fluctuations using functional integral methods
\cite{mohit2} and focusing on $T \geq T_c$. A holon pair field
is written as a two-component vector field: $F_{ij}(\tau) = F_{i,\eta}(\tau)$, 
where $\eta = (x,y)$. Let
$F_{i\eta}(\tau) = \frac{1}{\sqrt{N\beta}}\sum _{p} e^{i\vk.(\bf r_i + \bf\eta/2) - i\omega\tau}F_{\eta}(\vk,\omega),$ 
where $\bf\eta$ is the unit vector, and $p = (\vk,\omega)$. 
The Hamiltonian density is given by
\beq H = \sum _p\xi (\vk)h^*_ph_p - \sum _{p\eta} E_{\eta}(\vk) F^*_{\eta}(p)F_{\eta}(p), \eeq
where $\xi (\vk) = \epsilon_h(\vk) - \mu_h$, and
\beq E_{x,y}(\vk) = 2t_0 \cos k_{y,x} + 2t_{z0}\cos k_z,\eeq
is the hopping energy for a holon pair. Here
$t_0 = t_sA^2$, and $t_{z0} = t_{zs}A^2$. 
We introduce an order-parameter field $\Delta _{\eta} (p)$, and 
do a Hubbard-Stratonovich transformation to obtain the action
$S = S_0 - \sum _{p\eta}\lbrack \Delta^*_{\eta}(p)\Delta _{\eta}(p) - \sqrt E_{\eta}(\vk)\Delta^*_{\eta}(\vk)F_{\eta}(p) + c. c. \rbrack$, 
where $S_0$ is the free (quadratic) holon part. Integrating out the holons and keeping terms to second order in $\Delta$
we obtain the effective action describing pairing fluctuations:
\beq S_{fl} = - \sum _{p\eta\rho} \lbrack \delta _{\eta\rho} - (E_{\eta}(\vk)E_{\rho}(\vk))^{1/2}\Pi_{\eta\rho}(p)\rbrack \Delta^*_{\eta}(p)\Delta _{\rho}(p), \eeq 
where $\Pi_{\eta\rho}(p) = <F^*_{\rho}(p)F_{\eta}(p)>$
is the pair correlation function for noninteracting holons, which  is given by
\begin{eqnarray}
\Pi _{\eta\rho}(\vq,\omega) & = & - \frac{1}{N}\sum _{\vk}\frac{\sin k_{\eta}\sin k_{\rho}}
{i\omega - \xi (\vk + \vq/2) - \xi(\vk - \vq/2)}\nonumber \\
& & \lbrack \tanh \frac{\xi (\vk + \vq/2)}{2T}
+ \tanh \frac{\xi (\vk - \vq/2)}{2T}\rbrack.\nonumber \\
& & 
\end{eqnarray} 
The sum is over the 2d Brillouin zone and $k_B = 1$. 

The thermodynamic potential is given by 
$\Omega = \Omega _1 + \Omega _2$, where $\Omega _1$ is the free holon part, and 
$\Omega _2 = T\sum _p \ln \lbrack(1 - \lambda _+(p))(1 - \lambda _-(p))\rbrack,$ 
is the fluctuation contribution, with 
 \begin{eqnarray}
\lambda _{\pm} & = & \frac{1}{2}\lbrack E_x\Pi _{xx} + E_y\Pi _{yy}
\pm \lbrace (E_x\Pi _{xx}
  - E_y\Pi _{yy})^2 \nonumber\\ 
 & & + 4E_xE_y\Pi _{xy}\Pi_{yx}\rbrace ^{1/2}\rbrack.
\end{eqnarray}
Since $E_x(0) = E_y(0) = 2t_0 + 2t_{z0}$, we obtain, for $p = 0$ (noting that $\Pi _{xy} < 0$)
$\lambda _{\pm} = E_x(0)\lbrack \Pi _{xx}(0) \mp \Pi _{xy}(0)\rbrack. $
The order-parameter equation is given by $1 = \lambda _+(0)$, 
the larger eigenvalue. At $T_c$ it reads
\beq \frac{1}{t_0 + t_{z0}} = \frac{1}{N}\sum _{\vk} (\sin k_x - \sin k_y)^2
\frac{\tanh (\xi(\vk)/2T_c)}{\xi(\vk)}. \eeq
This corresponds to a $d$ wave.  
The holon density is given by $x = x_1 + x_2$, where 
\beq x_1  =  \frac{1}{2N} \sum _{\vk} \lbrack 1 - \tanh (\xi(\vk)/2T)\rbrack \eeq 
is the contribution from free holons, and 
\beq x_2 =  - N^{-1}\partial \Omega _2/\partial \mu_h \eeq
is the fluctuation contribution.

Solving Eqs. (11-13) we obtain $\mu _h$ and $T_c$ for small $x$. 
We take the bottom of the holon band to be at zero. In the MF (BCS)
approximation fluctuations are neglected ($x = x_1$). The MF 
$T_c$, which we denote by $T_p$, is a measure of the pair binding energy, 
and remains finite as $x \rightarrow 0$, as shown in in Fig. (1). 
 When fluctuations are included, 
real-space pair states appear as poles of the two-holon Green's 
function, i.e., as zeroes of $Re(1 - \lambda_{\pm}(\vq,\omega)$,
which are at $\omega = \omega _{\pm}(\vq) = E_{p,\pm}(\vq) - \mu_p$.
(We have made an analytic continuation to real frequencies).
Here,  $E_{p,\pm}$ is the energy of the pair, $\mu _p$ is the pair
chemical potential, such that $\omega _{\pm}(\vq) \ge 0$. We need to 
consider only the lower pair band $E_{p+}(\vq)$ which has
states below the free holon-holon ($h$-$h$) continuum, therefore are long lived. 
Once formed, the states near the bottom of the band (which is at $\vq = 0$),
are quickly occupied by holons so that $x_1$ is negligible, and $x \approx x_2$. 
Then, $\mu _h \sim E_{p+}(0)/2$ is negative, and binding energy is $\sim 2|\mu _h|$.
\begin{figure}[htbp]
\centering
\includegraphics[height=6cm,width=8.0cm,angle=0]{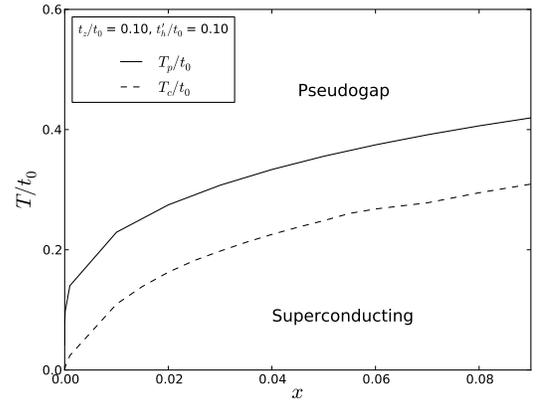}
\caption{Critical temperature $T_c$ and the holon pair dissociation
temperature $T_p$ as a function of hole density. In the region, $T_p > T > T_c$,
pairs coexist with holons.}
\end{figure}  
Bose condensation occurs for $\omega = \omega _+(\vq = 0) = 0$, so that  
$\mu _p = E_{p+}(0) \sim 2|\mu_h|$. Numerical solution shows that, for small $x$,
$|\mu _h| \sim t_s >> T_c$, so that pairs are strongly bound. Since pairs are bosons we need to
consider only states near $\vq = 0$. It is easily shown that $\lambda _{\pm}$ is symmetric 
in the $q_x,q_y$ plane. Then for small $\vq$ the pair spectrum has the form
$$\omega _+(\vq) = a(q_x^2 + q_y^2) + b q_z^2, $$
which gives
$$T_c \approx 4.66 (x_2)^{2/3}(a^2b)^{1/3}.$$
The parameters $a,b$ also depend on $T_c$, and indirectly on $x$, and 
obviously $x_1 = x - x_2$ is not zero. However, these corrections arise from
free holons, which require a finite energy $= 2|\mu_h|$ to produce. For small
$T_c/|\mu _h|$, we obtain
$x_1 \approx e^{-|\mu _h|/T_c}Z_1$, where $Z_1 = \frac{1}{N}\sum _\vk e^{- \epsilon _h (\vk)/T_c}.$
We find from numerical calculations that $a \sim t_0$, and $b$ scales is $t_{z0} << t_0$.
Since, $T_c$ also depends on $x_2 \sim x$, 
$T_c/\mu_h \sim (x^2t_{z0}/t_0)^{1/3}$ is small, and $x_1$ vanishes exponentially.
Fig. 1 shows $T_c$ and $T_p$ as a function of $x$.

The bound pairs continue to exist up to $T_p > T_c$. Holons are gapped below $T_p$,
but there is no sharp transition. Now, a mobile holon pair is {\em hidden}, 
as it is not gauge invariant. However, in the present case, 
it becomes observable through the gauge-invariant physical electron (singlet) 
pair represented by:  ${\cal C} _{ij} = -{\rm A}_{ij}{\rm F}_{ij}^{\dag}$. 
Ordinarily the electron pair Green's function 
$G^{el-pair}_{\eta\eta^{\prime}}(p) = - <{\cal C} _{\eta}(p){\cal C}^* _{\eta^{\prime}}(p)>$
is incoherent since it is a convolution. However, a coherent part emerges
as the spinon singlets condense below $T^*$. Replacing ${\rm A}_{ij}$ by its mean-field value, we 
obtain 
$G^{el-pair}_{\eta\eta^{\prime}}(p) = A^2G^F_{\eta\eta^{\prime}}(-p)$, 
where $G^F_{\eta\eta^{\prime}}(p) = - <F_{\eta}(p)F_{\eta^{\prime}}^*(p)>$.
To determine $G^{F}$ we use the exact relation 
$(E_{\eta}E_{\eta^{\prime}})^{1/2}<F^*_{\eta}F_{\eta^{\prime}}> = - \delta _{\eta\eta^{\prime}}
+ <\Delta^* _{\eta}\Delta _{\eta^{\prime}}>$. This leads to
\beq G^{el-pair}_{\eta\eta^{\prime}}(p) = 
- \frac{A^2Z_{\eta\eta^{\prime}}(-\vk)}{i\omega + \omega _+(-\vk)} + incoherent~ part,\eeq
which is a key prediction. Here
$$Z_{\eta\eta^{\prime}} = \frac{u^*_+(\eta^{\prime})u_+(\eta)}
{(E_{\eta}E_{\eta^{\prime}})^{1/2}(\partial\lambda _+/\partial\omega)} $$
evaluated at $\omega = \omega (\vk)$; $u_{\pm}(\eta)$ are the eigenvectors which diagonalize 
the effective action [Eq. 8], yielding eignevalues $\lambda _{\pm}$.
Eq. (14) describes a well-defined physical (gauge-invariant) pair excitation. However, unlike
a Cooper pair, it is observable above and below $T_c$, and repesents both electrons and holes. 
Since the pole is at negative energy $-\omega_+(-\vk)$, the actual excitation represents a
mobile hole pair of momentum $\vk$ and energy $\omega_+(\vk)$, of charge $2e$,
The spin part consists of a  spinon singlet (representing a localized electron pair). 
These are already condensed below $T^*$, and tend to overlap because of high density
($1 - x$). With increasing $x$ and/or $T$, the excitation will be less sharp due to broadening by 
phase (gauge) fluctuations associated with $A_{ij}$, and the factor $A^2$, 
which vanishes for $T \ge T^*$.

These results resolve a number of puzzling issues. Holon pairing implies that $\rho _s(0) \sim x$. 
Superconductivity by BEC explains the decline of $T_c$ with decreasing $x$. 
That $T_c$ does not vanish at some $x_{SC} \sim x_c$ is not suprising since the  renormalized Hamiltonian is 
not valid near $x_c$, where the spin gap is small, AF correlations are longer ranged and 
compete with superconductivity. Experimentally, $T_c \propto \rho_s ^{\gamma}$, i.e., both
vanish at same $x_{Sc}$, as if the effective carrier density is $x-x_{SC}$. Early reports indicated
that $\gamma = 1$ \cite{uem}; however, recently $\gamma = 0.61$ has also been reported \cite{hardy}, which
is not far form $2/3$. The existence of charged bound pairs above $T_c$ has been shown to
account \cite{alex} for the observed diamagnetism \cite{ong1}. On the other hand, since
the spin part of the pair is condensed below $T^* > T_c$, and the associated phase (gauge) 
fluctuations couple to holon pairs, vortex type excitations are likely to exist. 
These may account for the observed Nernst effect \cite{ong2},
although more work would be needed to sort out these issues. Interestingly, inthis theory,
spinon singlet condensation is two-dimensional,
so that the transition at $T^*$ is probably Kosterlitz-Thouless type, 
hence the lack of singularities. Then phase fluctuations are also 
two-dimensional. The theory is consistent with the observed 
decoupling between spin and charge 
responses. As $T$ decreases, the paramagnetic susceptibility decrease
starting at the spin-gap temperature $T^0$, and seemingly 
unaffected by charge pairing below $T_p$ and superconductivity at $T_c$ \cite{tim,sar10}.
Finally, though not considered here, nodal electrons can appear as collective excitations 
\cite{sar12}, but would not directly take part in pairing.

\end{document}